\documentclass[12pt]{article}

\pdfoutput=1

\usepackage{a4wide}
\usepackage[pdftex,usenames,dvipsnames]{color}
\usepackage{graphics}
\usepackage{amsfonts}
\usepackage{amssymb}

\newcommand{\nit}{\noindent}

\newcommand{\np}{\newpage}
\newcommand{\dsp}{\displaystyle}
\newcommand{\vs}[1]{\vspace{#1 ex}}
\newcommand{\hs}[1]{\hspace{#1 em}}
\newcommand{\bfr}{\begin{flushright}}
\newcommand{\efr}{\end{flushright}}
\newcommand{\bc}{\begin{center}}
\newcommand{\ec}{\end{center}}
\newcommand{\ben}{\begin{enumerate}}
\newcommand{\een}{\end{enumerate}}

\newcommand{\be}{\begin{equation}}
\newcommand{\ee}{\end{equation}}
\newcommand{\ba}{\begin{array}}
\newcommand{\ea}{\end{array}}
\newcommand{\ct}{\cite}
\newcommand{\bit}{\bibitem}

\newcommand{\ag}{\alpha}
\newcommand{\bg}{\beta}
\newcommand{\gam}{\gamma}
\newcommand{\del}{\delta}

\newcommand{\ve}{\varepsilon}

\newcommand{\thg}{\theta}
\newcommand{\kg}{\kappa}
\newcommand{\lb}{\lambda}
\newcommand{\sg}{\sigma}
\newcommand{\rg}{\rho}

\newcommand{\vf}{\varphi}
\newcommand{\og}{\omega}
\newcommand{\Gam}{\Gamma}

\newcommand{\Fg}{\Phi}

\newcommand{\Og}{\Omega}

\newcommand{\bfB}{\bold {B}}

\newcommand{\tog}{\tilde{\og}}

\newcommand{\cO}{{\cal O}}

\newcommand{\lh}{\left(}
\newcommand{\rh}{\right)}
\newcommand{\ld}{\left.}

\newcommand{\cotan}{\mbox{cotg\,}}

\newcommand{\der}{\partial}

\begin{document}

\pagestyle{empty}

\bfr
NIKHEF 2024-009
\efr

\vs{2}
\bc
{\bf \large Polar magnetic fields in black-hole space-times}
\vs{5} 

{\large R.\ Kerner} 
\vs{2}

Sorbonne University, Paris
\vs{2}

{\large G.\ Koekoek}
\vs{2}

Department of Gravitational Waves and Fundamental Physics, Maastricht University, \newline
and Nikhef
\vs{2}

{\large J.A.\ Schuring}
\vs{2}

Department of Gravitational Waves and Fundamental Physics, Maastricht University, \newline
and Nikhef
\vs{2} 

{\large and}
\vs{2}

{\large J.W.\ van Holten}
\vs{2}

Nikhef, Amsterdam and Lorentz Institute, Leiden University 
\vs{3}

June 21, 2024
\vs{5}
\ec

\nit
{\small {\bf Abstract} \\
To model magnetic fields of compact objects we solve the Maxwell equations in the background 
of the exterior static Schwarzschild and slowly rotating Kerr space-times. We impose the 
boundary condition that the electromagnetic fields are to vanish at infinity. A full set of solutions 
is obtained, describing axially symmetric magnetic fields, supplemented by axial electric 
fields in the case of non-vanishing rotation of the gravitational background. We study the 
motion of charged test particles in these combined gravitational and electromagnetic 
fields, in particular considering the conditions for circular equatorial orbits. Such orbits 
always exist in odd-multipole magnetic fields, and they can exist for particular radii 
in a combination of two or more even-multipole magnetic fields. Combinations of several 
odd-multipole fields can give rise to radial variation in the field orientation and the direction 
of motion of charged particles. Deviations from circularity are described using a perturbative 
approach. This also allows to study the stability of the parent circular orbits.  
}

\np

\pagestyle{plain}
\pagenumbering{arabic}

\section{Introduction \label{s1}}

Compact astrophysical objects are frequently accompanied by extended magnetic fields. 
These fields can have their origin in the object itself, as happens with white dwarfs or neutron 
stars, or they are caused by external currents of charged matter like accretion disks, as is 
particularly relevant for black holes \ct{eht-M87:2021,eht-sgrA:2022}. 

Configurations of magnetic fields in the background of Schwarzschild or Kerr geometries have 
been studied by many authors \ct{wald:1974}-\ct{karas:2023}; such an approximate description 
of the solutions of the Einstein-Maxwell equations are relevant when the back reaction of the 
external electromagnetic field on the space-time geometry can be neglected, i.e.\ when the 
effect of the energy density in this field on the space-time curvature is sufficiently small, as 
is often the case. These weak fields are sometimes refered to as test fields. 

In this paper we review some of the earlier results and then proceed to study the motion 
of charged particles in the combined gravitational and electromagnetic background, first of 
all in a Schwarzschild background. Charged particle motion, in particular on circular orbits, 
has been widely studied \ct{prasanna:1980}-\ct{baker:2023}. Here we extend the analysis 
by considering in detail the conditions for their stability, and by allowing non-circular 
orbits as well. In addition we show that combinations of different multipole fields allow 
discrete orbits that would not exist in only a dipole or single higher-order multipole field 
by itself. 

After considering magnetic fields in the static Schwarzschild geometry we also address weakly 
rotating compact objects in the context of the linear approximation of a Kerr background.
In that case in the frame of a static observer at asymptotically large distance any static 
axisymmetric magnetic field is necessarily accompanied by radial and polar electric fields. 
We conclude by summarizing our results; many details of our calculations are collected in
the appendices. Throughout this paper results are expressed in natural units in which the 
speed of light $c = 1$.

\section{Static magnetic fields in Schwarzschild space-time \label{s2}}

\nit
Full analytic expressions for static and axially symmetric magnetic test fields in  
Schwarz\-schild space-time have been derived by several authors. The precise 
form depends on the boundary conditions. A particular solution which has been
much studied is expressed in terms of the Killing vectors of the underlying static 
and spherically symmetric space  \ct{wald:1974}; magnetic test fields of this type 
are asymptotically uniform and finite. 

In the present paper we consider test fields which are localized around stars and 
black holes and vanish at asymptotically large distances; their structure close to 
the surface or horizon can still be allowed to vary, depending on the source of the 
magnetic field. In the case of a static and spherically symmetric space-time, employing 
standard Schwarzschild-Droste co-ordinates (reviewed in appendix \ref{a1}), one can 
fix a gauge such that the electro-magnetic vector potential has only one non-vanishing 
component $A_{\vf}$, satisfying the equation \ct{petterson:1974}-\ct{preti:2004}
\be
r^2 \der_r \left[ \lh 1 - \frac{2GM}{r} \rh \der_r A_{\vf} \right] + 
 \sin \thg \der_{\thg} \left[ \frac{1}{\sin \thg} \der_{\thg} A_{\vf} \right] = - r^4 \sin^2 \thg\, j^{\vf},
\label{n2.1}
\ee
where $j^{\vf}$ is the density of a current circulating the black hole and acting as 
a source for the magnetic field. Details of the derivation are given in appendix \ref{na2}. 

In regions where the current density vanishes: $j^{\vf} = 0$, the equation becomes separable;
taking a potential defined by a product
\be
A_{\vf} = f(r) \Fg(\cos \thg),
\label{n2.2}
\ee
the factors are related by coupled ordinary differential equations in $r$ and $x = \cos \thg$:
\be
r^2 \frac{d}{dr} \left[ \lh 1 - \frac{2GM}{r} \rh \frac{df}{dr} \right] = \lb f, \hs{1} \mbox{and} \hs{1}
(1 - x^2)\, \frac{d^2 \Fg}{dx^2} = - \lb \Fg,
\label{n2.3}
\ee
where $\lb$ is a constant common eigenvalue of the operators acting on $\Fg(x)$ and $f(r)$.
For the angular component $\Fg(x)$ there exists a complete set of eigenfunctions 
$\Fg_l(x)$ defined in terms of the Legendre polynomials $P_l(x)$, $l = 0,1,2,...$, by
\be
\Fg_l(x) = (1 - x^2)\, \frac{dP_l}{dx} = l \lh P_{l-1}(x) - x P_l(x) \rh.
\label{n2.4}
\ee
For $l = 0$ this implies that $\Fg_0(x) = 0$, therefore in the spectrum of $\Fg_l$
the value $l = 0$ may be disregarded. Note that, as the Legendre polynomials are solutions 
of the equations 
\be
\frac{d}{dx} \left[ \lh 1 - x^2 \rh \frac{dP_l}{dx} \right] = - l (l+1)\, P_l(x),
\label{n2.5}
\ee
it follows that 
\be
\frac{d\Fg_l}{dx} = - l (l+1) P_l(x). 
\label{n2.6}
\ee
Combining the above definitions and relations one then gets
\be
(1 - x^2)\, \frac{d^2\Fg_l}{dx^2} = - l (l+1) \Fg_l(x).
\label{n2.7}
\ee
Thus we obtain a complete set of solutions for the second equation (\ref{n2.3}) with eigenvalues 
\be
\lb = l (l+1), \hs{2} l = 1,2,...
\label{n2.8}
\ee
It then remains to solve for the corresponding radial functions $f_l(r)$ to obtain the 
general form of the vector potential in the form of a series:
\be
A_{\vf}(r,\thg) = \sum_{l = l}^{\infty} f_l(r) \Fg_l(\cos \thg).
\label{n2.9}
\ee
The radial functions satisfy the eigenvalue equation (\ref{n2.3}) with the eigenvalues 
(\ref{n2.8}). This equation has two kinds of solutions \ct{petterson:1974}; first there are 
solutions expressed in terms of an infinite series in powers of $1/r$: 
\be
f_l(r) = \sum_{n = l}^{\infty} c_n^{(l)} \lh \frac{2GM}{r} \rh^n,
\label{n2.10}
\ee
with the coefficients related to the first one  by 
\be
c^{(l)}_{l+k} = \frac{1}{k!} \prod_{m = 1}^{k} \lh \frac{(l+m)^2 - 1}{2l + m +1} \rh c^{(l)}_l, \hs{2}
 k = 1,2,... 
\label{n2.11}
\ee
These solutions are therefore defined by a single free normalization parameter $c_l^{(l)}$. 
As by construction they vanish in the limit $r \rightarrow \infty$, they are relevant 
especially --but not exclusively-- in the large-$r$ region. 

In addition to the infinite series solutions vanishing at infinity there exist polynomial solutions 
\be
A_{\vf}(r,\thg) = \sum_{l = l}^{\infty} g_l(r) \Fg_l(\cos \thg),
\label{n2.9a}
\ee
with
\be
g_l(r) = \sum_{n=2}^{l+1} a_n^{(l)} \lh \frac{r}{2GM} \rh^n.
\label{n2.12}
\ee
Here $a_2^{(l)}$ is a free parameter, in terms of which the other coefficients are
given by
\be
a^{(l)}_{2+k} =  \prod_{m=1}^k \lh \frac{m(m+1) - l(l+1)}{m(m+2)} \rh  a^{(l)}_2, 
  \hs{1} k = 1,...,l-1.
\label{n2.13}
\ee
Obviously these solutions do not vanish asymptotically for large $r$; therefore they can 
be relevant at most in a finite domain of $r$-values in the inner region of the magnetic 
field.  Explicit expressions for the first few functions $\Fg_l(\cos \thg)$, $f_l(r)$ and 
$g_l(r)$ with $l = 1, ..., 4$ are presented in the appendices \ref{na3} and \ref{na4}.

In the following we focus on the series solutions which vanish at infinity. From the 
vector potentials (\ref{n2.9}) one derives two non-vanishing components of the 
Maxwell tensor:
\be
F_{r \vf} = \sum_{l=1}^{\infty} f_l'(r) \Fg_l(\cos \thg), \hs{2} 
F_{\thg\vf} =  \sin \thg\, \sum_{l=1}^{\infty} l(l+1)\, f_l(r) P_l(\cos \thg).
\label{n2.14}
\ee
The corresponding components of the magnetic field strength $B_i = \tilde{F}_{0i}$, 
as defined in appendix \ref{na2}, are 
\be
B_r = \frac{1}{r^2} \sum_{l=1}^{\infty} l(l+1)\, f_l(r) P_l(\cos \thg), \hs{1}
B_{\thg} = - \frac{1}{\sin \thg} \lh 1 - \frac{2GM}{r} \rh \sum_{l=1}^{\infty} f_l'(r) \Fg_l(\cos \thg),
\label{2.20a}
\ee
whilst the azimuthal component vanishes: $B_{\vf} = 0$. \\

\nit
{\bf Dipole fields} \\
The lowest and often dominant component of the magnetic field is of dipole form
with $l = 1$; in that case 
\be
P_1(\cos \thg) = \cos \thg, \hs{2} \Fg_1(\cos \thg ) = 1 - \cos^2 \thg, 
\label{2.25}
\ee
and 
\begin{eqnarray}
f_1(r) &=& c_1^{(1)}\, \frac{2GM}{r} \lh 1 + \frac{3}{2} \frac{GM}{r} + 
  \frac{12}{5} \lh \frac{GM}{r} \rh^2 + ... \rh \nonumber \\
 f'_1(r) &=& - c_1^{(1)}\, \frac{2GM}{r^2} \lh 1 + 3\, \frac{GM}{r} + 
 \frac{36}{5} \lh \frac{GM}{r} \rh^2 + ... \rh. 
 \label{fprime}
\end{eqnarray}
These expressions can actually be written in closed form as
\be
\ba{l}
\dsp{ f_1(r) = - \frac{3 c_1^{(1)}}{2} \lh \frac{r^2}{2G^2M^2} \ln \lh 1 - \frac{2GM}{r} \rh 
 + \frac{r}{GM} + 1 \rh, }\\
 \\
\dsp{ f'_1(r) = - \frac{3c_1^{(1)}}{2GM} \lh \frac{r}{GM} \ln \lh 1 - \frac{2GM}{r} \rh + 
 \frac{1}{1 - \frac{2GM}{r}} + 1 \rh. }
\ea
\label{dipole}
\ee
The non-vanishing magnetic field components then have expansions
\be
B_r = \frac{2\mu}{r^3}\, \cos \thg \lh 1 + \frac{3GM}{2r} + ... \rh, \hs{2} 
B_{\thg} = \frac{\mu}{r^2}\, \sin \thg \lh 1 + \frac{GM}{r} + ... \rh,
\label{2.27}
\ee
with the magnetic dipole moment $\mu$ defined by
\be
\mu = 2 GM c_1^{(1)}. 
\label{2.28}
\ee
Note that the corresponding polynomial solution with $l = 1$ follows from 
\be
g_1(r) = a_2^{(1)} \lh \frac{r}{2GM} \rh^2, \hs{1} g'_1(r) = a^{(1)}_2 \, \frac{r}{2(GM)^2},
\label{2.30}
\ee
and gives rise to magnetic fields 
\be
B_r = B_z \cos \thg, \hs{2} B_{\thg} = - B_z \lh r - 2GM \rh \sin \thg,
\label{2.31}
\ee
where $B_z$ is the constant magnetic field strength on the positive $z$-axis:
\be
B_z = \frac{a_2^{(1)}}{2 (GM)^2}.
\label{2.32}
\ee
This magnetic field is purely transverse in the equatorial plane and purely radial along 
the $z$-axis. This is the asymptotically constant field found in ref.\ \ct{wald:1974} 
specialized to the Schwarzschild case. In view of the boundary conditions at infinity 
we will not consider this solution in the following.

\section{The motion of charged test particles \label{s3}}

The motion of massive test particles in combined gravitational and magnetic fields is decribed 
by a world line $\xi^{\mu}(\tau)$ with tangent vector $u^{\mu} = \dot{\xi}^{\mu}$, which are 
solutions of the covariant Lorentz force equation
\be
\dot{u}^{\mu} + \Gam_{\lb\nu}^{\;\;\;\mu}(\xi) u^{\lb} u^{\nu} = 
 \frac{q}{m}\, F^{\mu}_{\;\,\nu}(\xi) u^{\nu}.
\label{3.1}
\ee
Here the overdot denotes a proper-time derivative: $\dot{\xi} = d\xi/d\tau$; the forces are 
local, being dictated by the values of the Riemann-Christoffel connection and the Maxwell 
tensor at the location of the world line. Our conventions for the metric and components of 
the connection in a Schwarzschild space-time are given in appendix \ref{a1}. The Maxwell 
tensor $F_{\mu\nu}$ has non-zero components $F_{r\vf}$ and $F_{\thg\vf}$ given in
equation (\ref{n2.14}).

As both the background geometry and the magnetic field are static, and they share 
axial symmetry, the energy and the angular momentum component in the $z$-direction 
of test particles are conserved. Indeed it is straightforward to check, using polar 
co-ordinates $\xi^{\mu} = (t,r,\thg,\vf)$, that the following are constants of motion:
\be
\ve = \lh 1 - \frac{2GM}{r} \rh u^t, \hs{2} 
\ell = r^2 \sin^2 \thg\, u^{\vf} + \frac{q}{m}\,A_{\vf}.
\label{3.2}
\ee
In addition, by definition of the proper time there is a constraint
\be
g_{\mu\nu}(\xi) u^{\mu} u^{\nu} = -1.
\label{3.3}
\ee
Eliminating $u^t$ and $u^{\vf}$ in terms of the constants of motion (\ref{3.2}) 
this becomes
\be
u^{r\,2} + \lh 1 - \frac{2GM}{r} \rh r^2 u^{\thg\,2} = \ve^2 - \lh 1 - \frac{2GM}{r} \rh 
 \left[ 1 + \frac{1}{r^2 \sin^2 \thg} \lh \ell - \frac{q}{m}\, A_{\vf} \rh^2 \right].
\label{3.4}
\ee
The $\thg$-component of the world line equation (\ref{3.1}) can be written as 
\be
\frac{d}{d\tau} \lh r^2 u^\thg \rh = \sin \thg \cos \thg\, r^2 u^{\vf\,2} + 
 \frac{q}{m}\, F_{\thg\vf}\, u^{\vf},
\label{3.5}
\ee
Now recall, that for even $l = 2n$ the polynomials $P_l(\cos \thg)$ and $\Fg_l(\cos \thg)$ 
in the equatorial plane take the values
\be
P_{2n}(0) = \frac{(-1)^n}{2^{2n}} \frac{(2n)!}{(n!)^2} \hs{1} \mbox{and} \hs{2} \Fg_{2n}(0) = 0,
\label{3.5a}
\ee
whilst for odd $l = 2n+1$ 
\be
P_{2n+1}(0) = 0, \hs{2} 
\Fg_{2n+1}(0) = (2n+1) P_{2n}(0) = \frac{(-1)^n}{2^{2n}} \frac{(2n+1)!}{(n!)^2}.
\label{3.5b}
\ee
It follows that the individual contributions of the even-$l$ multipole fields to the magnetic field 
strength $F_{\thg\vf}$ in the equatorial plane never vanish, and charged particles in 
non-radial orbits experience a transverse Lorentz force. Indeed in the equatorial plane 
$\cos \thg = 0$, in a magnetic multipole field with $l = 2n$ 
\[ 
\frac{d}{d\tau} \lh r^2 u^{\thg} \rh_{l = 2n} = \frac{q \ell}{mr^2}\, 2n (2n+1) f_{2n}(r) P_{2n}(0).
\]
This vanishes only on radial trajectories, with $\ell = 0$. In contrast, for odd multipoles 
with $l = 2n+1$ the magnetic field in the equatorial plane is perpendicular to that 
plane, and the Lorentz-force component on the right-hand side of equation (\ref{3.5}) 
vanishes there. Planar equatorial orbits then exist in odd multipole fields with
\be
r^2 u^{\vf} = \ell - \frac{q}{m}\, (2n+1) P_{2n}(0)\, f_{2n+1}(r).
\label{3.6}
\ee
Note however, that although individual even-$l$ multipole fields do not allow for planar 
equatorial orbits, whenever there are several of these multipole fields they will in 
general cancel each other in the equatorial plane for specific values of the radial 
co-ordinate $r$, depending on the relative strengths as determined by the coefficients 
$c_l^{(l)}$; this is clear from the alternating signs in (\ref{3.5a}). As a consequence there
exist circular orbits at characteristic radius $r = R$ such that
\be
\sum_{n = 1}^{\infty} 2n (2n+1) P_{2n}(0) f_{2n}(R)= 0.
\label{3.7}
\ee
As an example $F_{\thg\vf}$ is plotted in fig.\ 3.1 as a function of $r/2GM$ for several 
combinations of quadrupole ($l=2$) and hexadecapole ($l=4$) fields. 
\bc
\vs{-2}
\scalebox{0.3}{\includegraphics{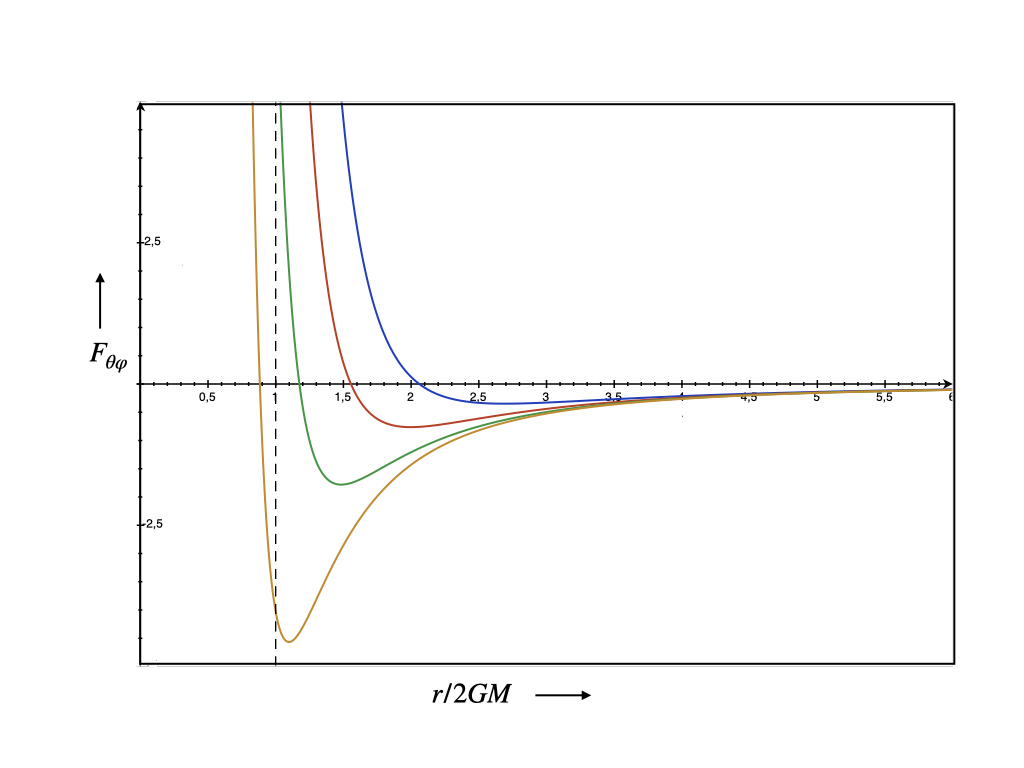}}
\vs{0}

{\footnotesize \begin{tabular}{ll} Fig.\ 3.1: & $F_{\thg\vf}$ as a function of $r/2GM$ in the 
 equatorial plane for combinations of quadrupole \\
 & and hexadecapole fields with relative strength 1 (blue), 1/2 (red), 1/4 (green) \\
 & and 1/8 (yellow); the last one has no zero-point outside the horizon.
 \end{tabular}}
\ec
Where $F_{\thg\vf} = 0$ a stable circular orbit exists, provided it lies outside the horizon 
$r/2GM = 1$ and the satisfies stability conditions to be discussed later on. 
\bc
\scalebox{0.3}{\includegraphics{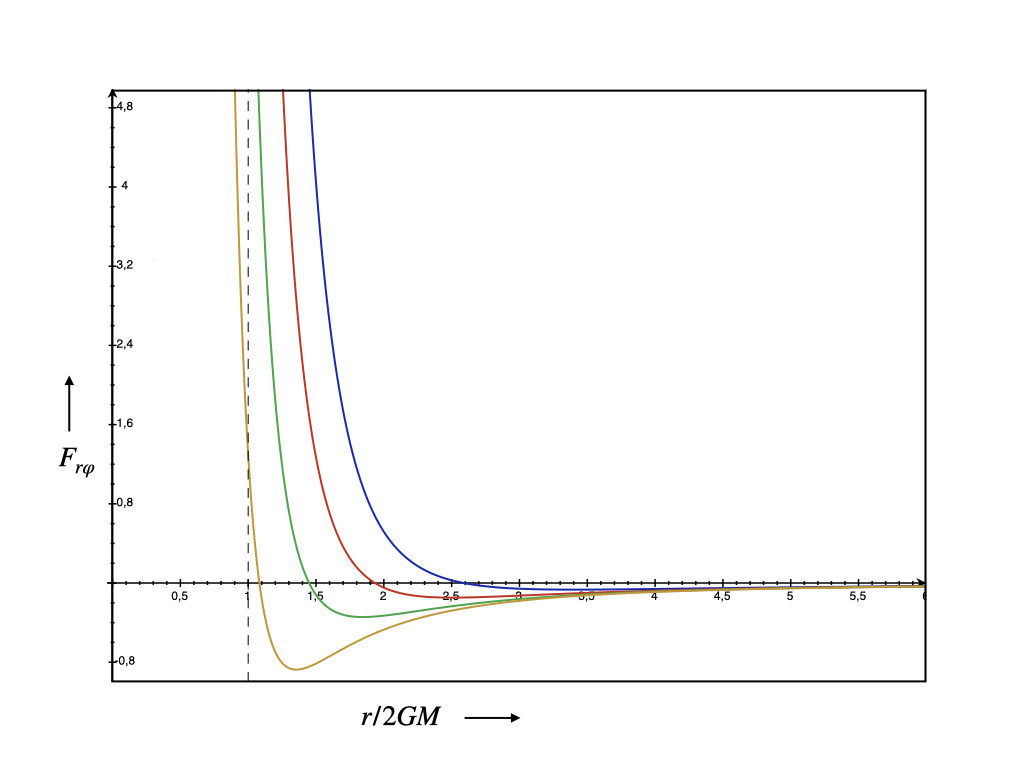}}
\vs{2}

{\footnotesize \begin{tabular}{ll} Fig.\ 3.2: & $F_{r\vf}$ as a function of $r/2GM$ in the 
 equatorial plane for combinations of dipole and \\
 & octopole fields with relative strength 1 (blue), 1/2 (red), 1/4 (green) and 1/8 (yellow).
 \end{tabular}}
\ec
\vs{1}

\nit
Fig.\ 3.2 shows $F_{r\vf}$ for similar combinations of dipole ($l = 1$) and octopole ($l=3$) 
fields. The absolute values depend on the scale set by $c_1^{(1,2)}$; however the zeros 
are not affected by the scale. These combinations of odd multipole fields allow for 
equatorial orbits of any radius outside the horizon meeting all stability criteria, but 
whenever $F_{r\vf} = 0$ the magnetic field and the Lorentz force on charges moving 
in the equatorial plane changes sign.

\section{Circular equatorial orbits  \label{s4}}

Generic circular orbits in the equatorial plane with fixed $r = R$ exist for dipole and 
other odd-$l$ multipole fields. On circular orbits both the radial velocity and the 
radial acceleration vanishes. They also require $F_{\thg\vf}(R,\pi/2) = 0$. Writing 
$u^{\vf} = \og_R$ the radial component of equation (\ref{3.1}) then implies 
\be
\lh 1 - \frac{2GM}{R} \rh \lh \frac{GM}{R^3}\, u^{t\,2} - \og_R^2 \rh = - \frac{qB_{\thg}}{mR}\, \og_R, 
\label{4.1}
\ee
with $B_{\thg}$ as defined in (\ref{2.20a}). In addition the constraint (\ref{3.3}) with 
$u^r = u^{\thg} = 0$ becomes:
\be
\lh 1 - \frac{2GM}{R} \rh u^{t\,2} = 1 + R^2 \og_R^2.
\label{4.2}
\ee
Combining these results provides a relation between the angular velocity and the 
orbital radius:
\be
\lh 1 - \frac{3GM}{R} \rh \og_R^2 - \frac{qB_{\thg}}{mR}\, \og_R = \frac{GM}{R^3}.
\label{4.3}
\ee
For all values of $R \neq 3GM$ this quadratic equation for $\og_R$ has general solutions
\be
R \og_R =  \frac{1}{1 - \frac{3GM}{R}} \lh \frac{qB_{\thg}}{2m}\, \pm  
 \sqrt{ \frac{GM}{R} \lh 1 - \frac{3GM}{R} \rh + \lh \frac{qB_{\thg}}{2m} \rh^2 } \rh. 
\label{4.3a}
\ee
However, provided $qB_{\thg} \neq 0$, for the special circular orbit at $R = 3GM$  
the equation reduces to a linear one, in which case
\be
R \og_R = - \frac{m}{3\, q B_{\thg}}, \hs{2} R = 3GM. 
\label{4.3d}
\ee
For electrically neutral particles, or in the absence of a magnetic field, this circular 
orbit is available only to massless particles like photons. A number of physically 
relevant solutions (\ref{4.3a}) have been plotted in fig.\ 4.1 for $R > 2 GM$.

\bc
\scalebox{0.32}{\includegraphics{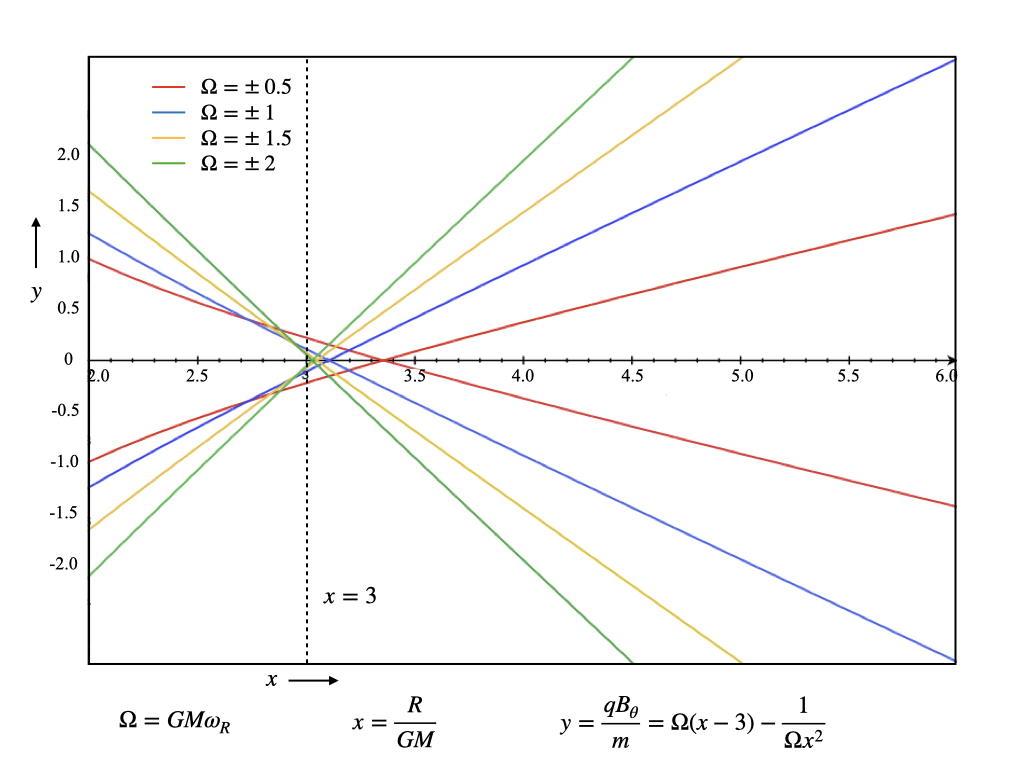}}
\vs{-0.5}

{\footnotesize Fig.\ 4.1: Solutions of equation (\ref{4.3a}).}
\ec
Solutions for test particles with opposite charge or opposite magnetic field strength, 
such that $qB_{\thg} \rightarrow - qB_{\thg}$, are related by opposite angular velocity: 
$\og_R \rightarrow - \og_R$. For each of these cases there is always a solution 
with $R > 3GM$. From eq.(\ref{4.3a}), it follows that a second solution with 
$2GM < R < 3 GM$ also exists provided 
\be
\lh \frac{qB_{\thg}}{m} \rh^2 \geq \lh \frac{2GM}{R} \rh^2 \lh 3 - \frac{R}{GM} \rh.
\label{4.4}
\ee
When this condition is not met, no circular orbit inside $3GM$ exists. 

Two extreme cases are $q = 0$ when only gravity acts: 
\be
\og^{\pm}_R = \pm \sqrt{\frac{GM}{R^3} \frac{1}{1 - \frac{3GM}{R}}};
\label{4.3b}
\ee
and $GM = 0$ when there is only a magnetic field causing cyclotron motion:
\be
\og^-_R = 0, \hs{2} \og^+_R = \frac{qB_{\thg}}{mR}. 
\label{4.3c}
\ee

\section{More on bound equatorial orbits \label{s5}}

Starting from the circular orbits discussed in the previous section, one can construct 
more general bound motions using the method of relativistic epicycles 
\ct{collistete:2001, collistete:2002}. 
In this procedure bound orbits are obtained as deformations of circular motion in a 
perturbative expansion. In this paper we discuss only first-order deviations from circular
motion; these deviations also provide information on the stability of circular orbits. It is 
straightforward to extend the procedure to higher orders \ct{koekoek:2011,vholten:2016}. 

The deviation of any bound orbit from a circular one is parametrized by 
$\xi^{\mu} = \xi^{\mu}_0 + n^{\mu}$, where to first order the deviation $n^{\mu}$ is a 
solution of the extended relativistic deviation equation
\be
\frac{D^2 n^{\mu}}{D\tau^2} = R_{\kg\nu\lb}^{\;\;\;\;\;\mu}\,  u^{\kg} u^{\lb} n^{\nu} 
 + \frac{q}{m}\, T^{\mu},
\label{5.1}
\ee
where
\be
T^{\mu} = F^{\mu}_{\;\,\nu} \frac{Dn^{\nu}}{D\tau} + n^{\nu} D_{\nu} F^{\mu}_{\;\,\lb} u^{\lb}.  
\label{5.0}
\ee
In these expressions the Riemann curvature $R_{\kg\nu\lb}^{\;\;\;\;\;\mu}(\xi_0)$, 
the magnetic field strength $F^{\mu}_{\;\,\nu}(\xi_0)$ and the proper velocity 
$u^{\mu} = \dot{\xi}^{\mu}_0$ are evaluated on the circular reference orbit \ct{balakin:2000}.  

We first consider deviations out of the equatorial plane, parametrized by $n^{\thg}$. 
As on a circular equatorial orbit $u^r = u^{\thg} = \cos \thg = 0$ and we require 
$F_{\thg\vf} = 0$ in the plane, it follows that the various contribution combine to
\be
\ba{lll}
\ddot{n}^{\thg} & = & \dsp{ - \frac{GM}{R^3} \left[ \lh 1 - \frac{2GM}{R} \rh u^{t\,2}
 + 2 R^2 u^{\vf\,2} \right] n^{\thg} + \frac{qu^{\vf}}{mR^2} \left[ R \lh 1 - \frac{2GM}{R} \rh F_{r\vf}
  + \der_{\thg} F_{\thg\vf} \right] n^{\thg} }\\
 & & \\
 & = & \dsp{ - \lh \og_R^2  + \frac{q\nu_R}{m}\, \og_R \rh n^{\thg}, \hs{1} \mbox{where} \hs{1}
  \nu_R = \ld \frac{dB_r(R)}{d\cos \thg} \right|_{\cos \thg = 0}. }
\ea
\label{5.2}
\ee
Clearly the orbital deviations in the plane do not mix with the resulting periodic transverse 
motion, and hence can be treated separately. The solutions of Eq.(\ref{5.2}) are periodic 
with frequency $\omega_\theta$ given by
\begin{equation}
\omega^2_\theta = \og^2_R \lh 1  + \frac{q\nu_R}{m \og_R} \rh,
\label{omegatheta}
\end{equation}
unless the squared frequency $\omega^2_\theta$ becomes negative, for 
$q\nu_R/m\og_R < -1$. This value represents the limit of stabilty of the orbit with respect 
to transverse fluctuations.

The remaining components of the deviation vector $n^{\mu}$ are coupled by Eq.(\ref{5.0}) and 
can be expressed in the form 
\be
\lh \ba{ccc} \frac{d^2}{d\tau^2} & \ag \frac{d}{d\tau} & 0 \\
                                                & & \\
            \bg \frac{d}{d\tau} & \frac{d^2}{d\tau^2} - \kg & - \gam \frac{d}{d\tau} \\
                                               & & \\
            0 & \eta \frac{d}{d\tau} & \frac{d^2}{d\tau^2} \ea \rh \lh \ba{c} n^t \\ \\ n^r \\ \\ n^{\vf} \ea \rh = 0,
\label{5.3}
\ee
where
\be 
\ba{l}
\dsp{ \ag = \frac{2GM}{R^2} \frac{u^t}{1 - \frac{2GM}{R}}, \hs{7.5}
 \bg = \frac{2GM}{R^2} \lh 1 - \frac{2GM}{R} \rh u^t, }\\
  \\
\dsp{ \gam = \lh 1 - \frac{2GM}{R} \rh 2Ru^{\vf} - \frac{qB_{\thg}}{m}, \hs{1.5} 
 \eta = \frac{2u^{\vf}}{R} - \frac{qB_{\thg}}{mR^2} \frac{1}{1 - \frac{2GM}{R}}, }\\
  \\
\dsp{ \kg =  3 \lh 1 - \frac{2GM}{R} \rh u^{\vf\,2} - \frac{2qB_{\thg}u^{\vf}}{mR}\, 
 \frac{R - 3GM}{R - 2GM} - \frac{q B'_{\thg}\, u^{\vf}}{m} }\\
  \\
\dsp{ \hs{2} =\, \frac{3GM}{R^3} \frac{R - 2GM}{R-3GM} + \frac{qB_{\thg}\, \og_R}{mR} 
 \frac{R^2 - 6G^2M^2}{(R-2GM)(R-3GM)} - \frac{qB'_{\thg}\, \og_R}{m}. }  
\ea
\label{5.4}
\ee
Here $B'_{\thg}(R) = (\der_r B_{\thg})(R)$. The eigenvalues $i \tog$ of the operator 
$d/d\tau$ in equation (\ref{5.3}) are solutions of the characteristic equation 
\be
\tog^4 \lh \tog^2 - \eta \gam + \ag \bg + \kg \rh = 0.
\label{5.5}
\ee
Apart from the zero-modes, the solutions are
\be
\ba{lll}
\tog^2 & = & \dsp{ \frac{GM}{R^3} \frac{R - 6 GM}{R - 3GM} - 
 \frac{qB_{\thg}\,\og_R}{mR} \frac{R^2 - 4GMR + 6G^2M^2}{(R-2GM)(R-3GM)}  }\\
 & & \\
 & & \dsp{ + \lh \frac{qB_{\thg}}{m} \rh^2 \frac{1}{R(R-2GM)} + \frac{qB_{\thg}'\,\og_R}{m} , }
\ea
\label{5.6}
\ee
resulting in periodic expressions for $n^t, n^r, n^\varphi$ for real frequencies $\tilde{\omega}$.

As in general the period of the deviations is different from that of the circular orbit, the 
periastron will perform a precession the phase of which is determined by the ratio of the 
angular frequencies; more precisely, with $n^r(\tau) = n^r(0) \cos \tog \tau$ it follows that 
the angles of two periastra at proper time $\tau = 0$ and $\tau_n$ differ by
\be
\tog \tau_n = n \pi \hs{1} \Rightarrow \hs{1} \vf(\tau_n) - \vf(0) = \frac{\og_R}{\tog}\, n \pi.
\label{5.7}
\ee
Following the criterion for transverse stability, orbital stability for fluctuations in the equatorial 
plane requires the squared frequency eq.(\ref{5.6}) to be non-negative, whereas exponentially 
run-away solutions are found when it is negative. Thus unstable orbits are possible for 
appropriate strength and sign of the magnetic interaction term $q B_\theta$. 

Specifying to a single-multipole magnetic field of rank $l$, eq.(\ref{2.20a}) implies
\begin{equation}
B'_{l\theta} = \left(\frac{2GM}{R^2}\frac{1}{1-\frac{2GM}{R}}+\frac{f_l''}{f'_l}\right)B_{l\theta}
\end{equation}
the orbital frequency $\omega_R$ is expressed in terms of the magnetic field $B_{l\theta}$, 
and the limit of planar stability $\tilde{\omega}^2 = 0$ becomes a quadratic equation for 
$B_{l\theta}$, which can easily be solved algebraically. 

\section{Kinematical stability of circular orbits for dipole fields \label{ns.6}}

We will now collect results on the existence and stability of circular orbits in a Schwarzschild 
space-time, applying them to the special case of a dipole field with $l=1$. We use the 
dimensionless quantities introduced in fig.\ 4.1;
\be
x = \frac{R}{GM}, \hs{2} \Og = GM \og_R, \hs{2} y = \frac{q B_{\thg}}{m}.
\label{n6.x}
\ee
As we consider the exterior of horizon, $x > 2$.
By eq.\ (\ref{2.27}) 
\be
y= \frac{q\mu}{m (GMx)^2}\, h(x),
\label{n6.0}
\ee
where using eq.\ (\ref{dipole})
\be
h(x) = \frac{3}{4}\left(1-\frac{2}{x} \right)x^2\, 
 \left( 1+\frac{1}{1-\frac{2}{x}}+x\, \ln \left( 1-\frac{2}{x}\right)\right). 
\ee
But eq.\ (\ref{4.3}) states that the dimensionless magnetic field expressed in terms 
of $x$ and $\Og$ is
\be
y(x, \Og) = (x-3)\, \Og - \frac{1}{x^2 \Og}. 
\label{n6.1}
\ee
Combining these results we find for $\Og(x)$ in terms of the physical data 
$(q,\mu,m,M)$ the expression
\begin{equation}
\frac{\Omega}{y} = \frac{1}{2(x-3)}\left(1 \pm \sqrt{1+\frac{4(x-3)}{x^2y^2}}\right).
\label{eq:OmegaOvery}
\end{equation}
Real solutions for $\Og$ exist iff
\be
y^2 + \frac{4}{x^2} \lh x - 3 \rh \geq 0.
\label{n6.1.a}
\ee
This inequality is automatically satisfied for all $y$ if $x > 3$. 
The circular motion is stable against fluctuations in the transverse direction (out of the plane)
provided $\og_{\thg}^2 > 0$; in view of eq.\ (\ref{omegatheta}) this requires 
\[
\frac{q\nu_R}{m\og_R} > -1 . 
\]
Now for $l = 1$, using eq.\ (\ref{n2.3})
\be
\ld \nu_R \right|_{l=1} = \frac{2\mu}{(GM x)^3}\, k(x), \hs{2} k(x) = 
  -\frac{3}{4}x\left( \frac{x^2}{2}\ln \left( 1-\frac{2}{x}\right)+x+1 \right).
\label{n6.0.a}
\ee
Therefore stability requires
\be
\frac{y}{\Og} > - \frac{xh(x)}{2k(x)}.
\label{n6.2}
\ee
This is always satisfied if $y/\Og > 0$. 
Combining this inequality with (\ref{n6.1}) it follows that 
\be
\Og^2 > \frac{2k(x)}{x^2 \left[2(x - 3)k(x) + x h(x)\right]}.
\label{n6.3}
\ee
We can also eliminate $\Og$ to get another inequality for $y$:
\be
\frac{4(x-3) k(x)}{x h(x)}\, \frac{1}{1 \pm \sqrt{1 + \frac{4(x-3)}{x^2y^2}}} > - 1, \hs{1}
 \frac{k(x)}{h(x)} =  1 + \frac{1}{2x} + \frac{7}{10x^2} + \frac{11}{10 x^3} + ...
\label{n6.4}
\ee
This inequality holds automatically if the l.h.s.\ is positive, which happens with 
the plus sign for the square root for all $y$ and $x > 3$.
Finally the circular motion must be stable under fluctuations in the equatorial plane,
which imposes the condition $\tog^2 > 0$. Now for a dipole field
\be
\frac{qB'_{\thg}}{m} = \frac{1}{GM} \frac{dy}{dx} = - \frac{2q\mu}{m (GMx)^3}\, k(x) 
 = -\frac{2k(x)}{R\,h(x)}\,y
\label{n6.5}
\ee
and hence stability under fluctuations in the equatorial plane follows via Eq.(\ref{5.6}) as
\be
1 + \frac{(x-2)(x-6)}{x^2 (x-3) y^2} - \frac{\Og}{y} \lh \frac{x^2 - 4x + 6}{x-3} + 
 2 (x-2) \frac{k(x)}{h(x)} \rh \geq 0.
\label{n6.7}
\ee
\nit
The results in equations (\ref{n6.1.a}), (\ref{n6.4}) and (\ref{n6.7}) provide implicit 
formulae for the curves dividing allowed stable regions from unstable regions in the 
$(x,y)$-plane. The results are shown in fig.\ 6.1, where the magnetic field strength 
is quantified in terms of the dipole moment $\mu$. The key features are the regions 
separated by the colored lines. The red line denotes the dividing line between 
kinematically allowed and disallowed orbits, where all of the region above the red 
line is allowed. The solid (dashed) blue line denotes the dividing line between planar 
stability and instability for orbits with minus (plus) sign in Eq.(\ref{eq:OmegaOvery}), 
where the region above (below) the blue line is stable. The solid (dashed) green line 
denotes the dividing line between transverse stability and instability for orbits with 
minus (plus) sign in Eq.(\ref{eq:OmegaOvery}), where for both signs the region right 
of the corresponding green line is stable. Analytical details are provided in appendix 
\ref{na6}. An important feature is that, in case of the minus sign, the smallest radius 
of equatorially stable circular orbits shifts from $R/GM = 6$ to $3$, while simultaneously 
the smallest radius of kinematically allowed circular orbit shifts from $R/GM = 3$ to the 
horizon of the black hole at $R/GM=2$. This is the result of gravitational attraction being 
counterbalanced by the Lorentz-force. \\
Note that in this section we have only considered the kinematical stability of circular 
orbits. In a complete treatment one would also have to consider the emission of 
electromagnetic and gravitational radiation, which ultimately destabilizes any 
classical bound orbit. 
\bc
\vs{-2}
\scalebox{0.4}{\hs{-7}\includegraphics{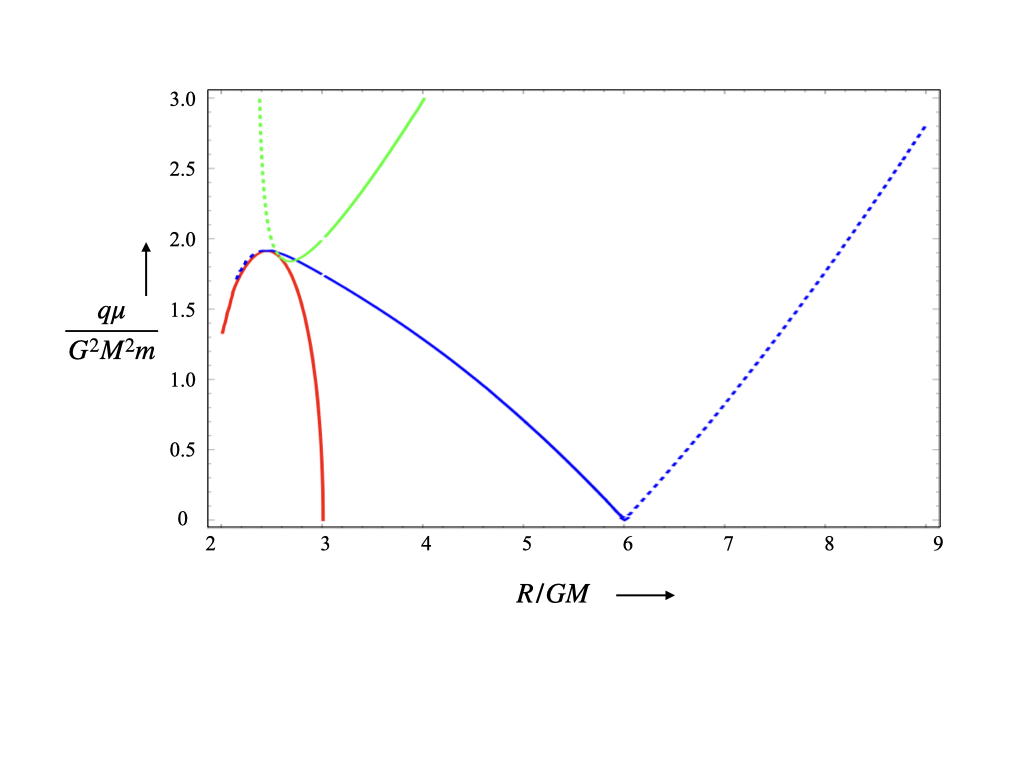}}
\vs{-14}
\ec

\nit
{\footnotesize Fig.\ 6.1: Regions of existence, and planar and perpendicular stabilities, for the orbits with plus and minus sign in Eq.(\ref{eq:OmegaOvery}).  The red line denotes the dividing line between kinematically allowed and disallowed orbits, the blue lines separates equatorial stability from instability, and the green lines separates transverse stability from instability. The resulting regions are described in the main text.}\\

\section{Magnetic fields of slowly rotating stars \label{s6}}

In the limit of slow rotation and distances $r \gg 2GM$ (that is, to first order in $G$) the line 
element of Kerr space-time takes the form of the Lense-Thirring metric
\be
ds^2 \simeq - \lh 1 - \frac{2GM}{r} \rh dt^2 + \lh 1 + \frac{2GM}{r} \rh dr^2 + r^2 d\thg^2 + 
 r^2 \sin^2 \thg\, d\vf^2 - \frac{4GJ}{r} \sin^2 \thg\, dt d\vf,
\label{6.1}
\ee
where $J = Ma$ is the angular momentum of the star. In this background space-time a 
static and axially symmetric Maxwell field with a gauge-fixed vector-potential of the form:
\[
A(r,\thg) = A_t(r,\thg) dt + A_{\vf}(r,\thg) d\vf,
\]
satisfies the curved-space Maxwell equations 
\be
\ba{l}
\dsp{ \left[ \der_r^2 + \frac{2}{r} \der_r + 
 \frac{1}{r^2 \sin \thg} \lh 1 + \frac{2GM}{r} \rh \der_{\thg} \sin \thg\, \der_{\thg} \right] A_t }\\
 \\
\dsp{ \hs{3} =\, - \frac{2GJ}{r^3} \left[ \der^2_r - \frac{1}{r}\, \der_r + \frac{1}{r^2\sin \thg} \der_{\thg} 
 \sin \thg \der_{\thg} \right] A_{\vf}, }\\
 \\
\dsp{ \der_r  \lh \frac{1}{\sin \thg} \lh 1 - \frac{2GM}{r} \rh \der_r A_{\vf} \rh + 
 \der_{\thg} \lh \frac{1}{r^2 \sin \thg} \der_{\thg} A_{\vf} \rh }\\
 \\
\dsp{ \hs{3} =\, \frac{2GJ \sin \thg}{r} \left[ \der_r^2 - \frac{1}{r} \der_r + 
 \frac{1}{r^2 \sin \thg} \der_{\thg} \sin \thg \der_{\thg} \right] A_t. }
\ea
\label{6.2}
\ee
Note that we cannot impose $A_t$ to vanish, unless $J = 0$. However, if we expand the 
solutions in powers of $G$, retaining only the terms to first order in $G$:
\be
A_t = a_t + G b_t, \hs{2} A_{\vf} = a_{\vf} + G b_{\vf},
\label{6.3}
\ee
and require the solutions to reduce to the corresponding Schwarzschild solution in the limit 
$J \rightarrow 0$, it follows that $a_t = 0$. As a result to this order the second equation (\ref{6.2})
reduces to the same equation for $A_{\vf}$ as in the Schwarzschild case, except that here 
we should only retain the terms up to order $G$:
\be
A_{\vf} = \sum_{l = 1}^{\infty} \frac{d_l}{r^{l}} \lh 1 + \frac{l(l+2)}{l+1}\, \frac{GM}{r} \rh \Fg_l(\cos\thg).
\label{6.3.a}
\ee
Inserting this expression into the right-hand side of the first equation, to the first order in $G$
only $a_{\vf}$ contributes:
\be
 \lh \der^2_r + \frac{2}{r}\, \der_r + \frac{1}{r^2 \sin \thg} \der_{\thg} \sin \thg\, \der_{\thg} \rh b_t
  = - \frac{2J}{r^3} \lh \der^2_r - \frac{1}{r}\, \der_r + \frac{1}{r^2 \sin \thg}\, \der_{\thg} \sin \thg\, \der_{\thg} \rh a_{\vf}.
\label{6.5}
\ee
The solution of this equation reads
\be
A_t = G b_t = - GJ \sum_{l=1}^{\infty} \frac{ld_l}{r^{l+3}}\, \cos \thg\, P_l(\cos \thg).
\label{6.7}
\ee
The coefficients $d_l$ determine the relative strengths of the multipole components. 
From these expressions we now deduce the components of the electromagnetic field 
strength tensor; the non-vanishing magnetic components are
\be
\ba{l}
\dsp{ F_{r\vf} = - \sum_{l = 1}^{\infty} \frac{ld_l}{r^{l+1}} \lh 1 + (l+2) \frac{GM}{r} \rh \Fg_l(\cos \thg), }\\
 \\
\dsp{ F_{\thg\vf} = \sum_{l \geq 1} \frac{l(l+1)d_l}{r^l} 
 \lh 1 + \frac{l(l+2)}{(l+1)}\, \frac{GM}{r} \rh \sin \thg\, P_l (\cos \thg), }
\ea
\label{6.8}
\ee
with corresponding non-vanishing electric components at order $G$:
\be
\ba{l}
\dsp{ F_{rt} = GJ \sum_{l = 1}^{\infty}  \frac{l(l+3) d_l}{r^{l+4}} \cos \thg\, P_l(\cos \thg), }\\
 \\
\dsp{ F_{\thg t} = GJ \sum_{l = 1}^{\infty} \frac{l d_l}{r^{l+3}} \sin \thg \lh P_l(\cos \thg) 
 + \cos \thg P'_l(\cos \thg) \rh. }
\ea
\label{6.9}
\ee
As the dependence on $J$ indicates, the appearance of these electric components is clearly 
a result of the rotation of the gravitational field in the presence of static magnetic field components. 
Following the analysis in appendix \ref{na2}, the $\thg$-component of the magnetic dipole 
field takes the standard form 
\[
B_{\thg} = \frac{\mu}{r^2}\, \sin \thg \lh 1 + {\cO}[G] \rh,
\]
after identifying $d_1 = \mu$. Of course, to first order in $G$ higher-$l$ multipoles are to be 
taken into account only as long as $d_{l+1} \gg \mu (GM)^l$. 

\section{Equatorial orbits in slowly rotating background \label{s7} }

As the metric (\ref{6.1}) is axially symmetric and stationary, the energy and $z$-component 
of angular momentum of test particles are again constants of motion:
\be
p_t = - m \ve, \hs{2} p_{\vf} = m \ell, 
\label{7.1}
\ee
with 
\be
\ba{l}
\dsp{ \ve = \lh 1 - \frac{2GM}{r} \rh u^t + \frac{2GJ}{r}\, \sin^2 \thg\, u^{\vf} - \frac{q}{m}\, A_t, }\\
 \\
\dsp{ \ell = r^2 \sin^2 \thg\, u^{\vf} - \frac{2GJ}{r}\, \sin^2 \thg\, u^t + \frac{q}{m}\, A_{\vf}. }
\ea
\label{7.2}
\ee
By inverting these equations we get the proper velocity components involved:
\be
\ba{l}
\dsp{ u^t = \lh 1 + \frac{2GM}{r} \rh \ve - \frac{2GJ\ell}{r^3} + 
 \frac{q}{m} \lh A_t + \frac{2GJ}{r^3}\, A_{\vf} \rh, }\\
 \\
\dsp{ u^{\vf} = \frac{1}{r^2 \sin^2\thg} \lh \ell - \frac{q}{m}\, A_{\vf} \rh + \frac{2GJ}{r^3}\, \ve, }
\ea
\label{7.3}
\ee
all expressions modulo terms $\cO[G^2]$, keeping in mind that according to eq.\ (\ref{6.7}) 
$A_t \sim \cO[G]$.

Next observe, that the electric fields (\ref{6.9}) vanish in the equatorial plane for odd multipoles; 
in fact for $l = 2n+1$ only the magnetic field component $F_{r\vf}$ remains in the plane 
$\thg = \pi/2$. Therefore in the presence of dipole and other odd-multipole fields there exist 
solutions of the equations of motion (\ref{3.1}) in the equatorial plane with $\sin \thg = 1$ and 
$u^{\thg} = 0$. On such orbits $A_t = 0$ and the hamiltonian constraint implies
\be
\ba{lll}
u^{r\,2} & = & \dsp{ \ve^2 - \lh 1 - \frac{2GM}{r} \rh \left[ 1 + \frac{1}{r^2} 
 \lh \ell - \frac{q}{m} A_{\vf} \rh^2 \right] - \frac{4GJ \ve}{r^3} \lh \ell - \frac{q}{m} A_{\vf} \rh 
 +  \cO[G^2].}
\ea
\label{7.4}
\ee 
For circular orbits with $r = R$ a constant, $u^{\vf} = \og_R$ a constant, 
$u^r = 0$ and $\dot{u}^r = 0$ it then follows that 
\be
\ba{l}
\dsp{ \ve^2 = \lh 1 - \frac{2GM}{R} \rh \lh 1 + R^2 \og_R^2 \rh, }\\
 \\
\dsp{ \lh 1 - \frac{3GM}{R} \rh \og^2_R - \left[ \frac{qB_{\thg}}{mR} - \frac{2GJ\ve}{R^3} \right] \og_R 
 = \frac{GM}{R^3}. }
\ea
\label{7.5}
\ee
For non-relativistic orbital velocities the last equation can be simplified (at order $G$) to give
\be
\frac{2GJ \ve}{R^3} \simeq \frac{2GJ}{R^3} + \frac{GJ\og_R^2}{R}. 
\label{7.6}
\ee
In this limit, and always up to terms of order $\cO[G^2]$:
\be
\lh 1 - \frac{3GM}{R} \rh \og^2_R - \lh \frac{qB_{\thg}}{mR} - \frac{2GJ}{R^3} \rh \og_R 
+ \frac{GJ \og_R^3}{R} = \frac{GM}{R^3}.
\label{7.7}
\ee
To check the stability of these orbits, we have again evaluated the world-line deviation 
equation (\ref{5.1}) for the deviations $n^{\thg}$ out of the equatorial plane; using the 
results of appendix \ref{na5} we get
\be
\ddot{n}^{\thg} = - \left[ \og^2_R - \frac{2GJ\ve}{R^3}\, \og_R - \frac{q}{mR^2} 
 \lh \der_{\thg} F_{\thg\vf} \og_R + \der_{\thg} F_{\thg t} \ve \rh \right] n^{\thg}.
\label{7.8}
\ee
In particular, for dipole fields:
\be
\ddot{n}^{\thg} = - \left[ \og_R^2 - \frac{2GJ\ve}{R^3}\, \og_R + 
 \frac{2q\mu \og_R}{mR^3} \lh 1 + \frac{3GM}{2R} \rh - \frac{2q\mu GJ \ve}{mR^6} \right] n^{\thg}.
\label{7.9}
\ee
This guarantees stability when 
\be
\frac{2GJ\ve}{R^3} < \og_R,
\label{7.10}
\ee
which is implicit in our approximation of a weakly rotating background geometry. 

\section{Discussion and summary \label{s8}}

The starting point of this paper is the development of a multipole expansion for 
magnetic fields in a background Schwarzschild space-time and its generalization 
to slowly rotating stationary curved space-times. These results are valid in the case 
the back reaction of the magnetic field on the space-time geometry can be neglected. 
This is correct when the space-time curvature is large compared to the energy density 
in the magnetic field. Now the leading term in the dipole magnetic field gives the energy 
density in the equatorial plane as 
\be
\frac{\ve_0}{2}\, \bfB^2 \simeq \frac{\ve_0 \mu^2}{4 r^6}, 
\label{8.1}
\ee
modulo higher-order corrections in $G$ falling off faster. Higher multipoles also fall 
of faster than the dipole field. In contrast the curvature of Schwarzschild space-time 
satisfies
\be
\sqrt{R_{\mu\nu\kg\lb} R^{\mu\nu\kg\lb}} = \frac{4\sqrt{3} GM}{r^3}.
\label{8.2}
\ee
Therefore the condition
\be
8 \pi G T_0^{\;0} \ll \sqrt{R_{\mu\nu\kg\lb} R^{\mu\nu\kg\lb}}, 
\label{8.3}
\ee
is satisfied everywhere if it is satisfied in the near-horizon region. 
Reinstating the speed of light, at the horizon the condition becomes 
\be
\frac{G \ve_0 \bfB^2}{c^2} \ll \lh \frac{c^2}{2GM} \rh^2.
\label{8.4}
\ee
For example, for neutron stars in the range of one to three solar masses this is satisfied 
for magnetic fields of $10^{15}$ T or less at the surface, which is about the observed 
upper limit. 

Once the fields are known one can establish if stable circular orbits in the equatorial plane 
for charged particles are possible. This requires the radial magnetic field component to 
vanish in this plane, a condition always satisfied by odd-$l$ multipole fields, like dipole 
and octopole fields. In the presence of even-$l$ multipole fields such orbits can only exist 
for special radii where the odd multipole fields happen to cancel each other. Orbits within 
the stable photon orbit $3GM$ are possible in principle, provided the magnetic field strength 
exceeds the limit (\ref{4.4}) and satisfies the stability conditions displayed in fig.\ 6.1. 

In addition to circular orbits, for the case of Schwarzschild space-time we have also found 
approximate non-circular orbits by the method of relativistic epicycles (world-line deviations). 
In general the periods of the epicycles differ from the parent orbital period, which leads 
to displacement of the periastron of the charged particles. Moreover, in the presence of
combinations of odd multipole fields the sign of the magnetic field can vary as a function 
of radial distance $r$, causing the direction of the rotation to change. 

We have derived conditions for existence and stability of circular orbits in the equatorial plane, 
in particular in non-rotating Schwarzschild space-time. We found that the radius of innermost 
circular orbit and the innermost stable circular orbit can shift, for appropriately high values of 
the magnetic interaction, towards the black hole horizon. For the special case of a dipole 
magnetic field in a Schwarzschild spacetime, we found the existence of a 'forbidden region' 
in between the horizon and the photosphere where circular orbits are not allowed to exist, 
and two regions of stability where the circular orbits are allowed and stable in planar and 
transverse directions in a specific range of magnetic field strengths. In principle, these 
regions are most likely to host charged particles in circular orbits, resulting in distinct 
gravitational wave and electromagnetic frequencies. We have not considered their 
dynamical stability, due to the emission of gravitational or electromagnetic radiation. We 
expect these regions to have measurable consequences for the detection of magnetic 
fields around black holes and determining their strengths. This remains to be addressed 
in future research.

We have not constructed orbits in the transverse direction, out of the equatorial plane. 
In general such orbits will have a helical structure, as the Lorentz force will causs charged 
particles to circle around magnetic field lines. We have considered the effect of slow rotation 
of the background gravitational field, by considering the first-order in $G$ approximation of 
the Kerr geometry. An interesting result is, that in that case also electric fields arise, both in 
the radial and the polar direction. However, again for odd-$l$ multipoles these electric field 
components vanish in the equatorial plane, still allowing for stable circular orbits. 

We have not considered here the origin of the magnetic fields. As is well-known white dwarfs 
and neutron stars can have their own intrinsic magnetic fields. For black holes the source of 
magnetic fields, as observed by the Event-Horizon Telescope, must be currents of charged 
particles such as accretion disks circulating close to the black hole itself. This presents an 
intricate problem of magnetohydrodynamics about which we presently have little to report.

\np
\appendix

\section{Schwarzschild geometry \label{a1}} 

In this appendix we summarize our conventions for the connection and curvature components 
of Schwarzschild geometry. Taking $c=1$ the line-element in standard Schwarzschild-Droste 
co-ordinates reads
\be
ds^2 = - \lh 1 - \frac{2GM}{r} \rh dt^2 + \frac{dr^2}{1 - \frac{2GM}{r}} + r^2 d\thg^2 + r^2 \sin^2 \thg\, d\vf^2.
\label{a.0}
\ee
The non-vanishing components of the corresponding Riemann-Christoffel connection are
\be
\ba{l}
\dsp{ \Gam_{tt}^{\;\;\;r} = \frac{GM}{r^2} \lh 1 - \frac{2GM}{r} \rh, \hs{3} 
\Gam_{rt}^{\;\;\;t} = \Gam_{tr}^{\;\;\;t} = \frac{GM}{r^2 \lh 1 - \frac{2GM}{r} \rh}, }\\
 \\
\dsp{ \Gam_{rr}^{\;\;\;r} = - \frac{GM}{r^2} \frac{1}{ 1 - \frac{2GM}{r}}, }\\
 \\
\dsp{ \Gam_{\thg\thg}^{\;\;\;r} = - r \lh 1 - \frac{2GM}{r} \rh, \hs{3.4} \Gam_{r\thg}^{\;\;\;\thg} = 
 \Gam_{\thg r}^{\;\;\;\thg} = \frac{1}{r}, }\\
 \\
\dsp{ \Gam_{\vf\vf}^{\;\;\;r} = - r \sin^2 \thg \lh 1 - \frac{2GM}{r} \rh, \hs{1} \Gam_{r\vf}^{\;\;\;\vf} = 
 \Gam_{\vf r}^{\;\;\;\vf} = \frac{1}{r}, }\\
 \\
\dsp{ \Gam_{\vf\vf}^{\;\;\;\thg} = - \sin \thg \cos \thg, \hs{3.9} \Gam_{\thg\vf}^{\;\;\;\vf} = \Gam_{\vf\thg}^{\;\;\;\vf} = \cotan \thg, }
\ea
\label{a.1}
\ee
whilst the non-vanishing components of the Riemann curvature tensor are given by
\be
\ba{l}
\dsp{ R_{trt}^{\;\;\;\;\;r} = \frac{2GM}{r^3} \lh 1 - \frac{2GM}{r} \rh, \hs{2.3} 
R_{rtr}^{\;\;\;\;\;t} = - \frac{2GM}{r^3 \lh 1 - \frac{2GM}{r} \rh}, }\\
 \\
\dsp{ R_{t\thg t}^{\;\;\;\;\;\thg} = - \frac{GM}{r^3} \lh 1 - \frac{2GM}{r} \rh, \hs{1.1} 
R_{\thg t\thg}^{\;\;\;\;\;t} = \frac{GM}{r}, }\\ 
 \\
\dsp{ R_{t\vf t}^{\;\;\;\;\;\vf} = - \frac{GM}{r^3} \lh 1 - \frac{2GM}{r} \rh, \hs{1} 
R_{\vf t\vf}^{\;\;\;\;\;t} = \frac{GM}{r}\, \sin^2 \thg, }
\ea
\label{a.2}
\ee
and
\be
\ba{l}
\dsp{ R_{r\thg r}^{\;\;\;\;\;\thg} = \frac{GM}{r^3 \lh 1 - \frac{2GM}{r} \rh}, \hs{1.6} 
R_{\thg r\thg}^{\;\;\;\;\;r} = \frac{GM}{r}, }\\
 \\
\dsp{ R_{r\vf r}^{\;\;\;\;\;\vf} = \frac{GM}{r^3 \lh 1 - \frac{2GM}{r} \rh}, \hs{1} 
R_{\vf r\vf}^{\;\;\;\;\;r} = \frac{GM}{r}\, \sin^2 \thg, }\\
 \\
\dsp{ R_{\thg\vf\thg}^{\;\;\;\;\;\vf} = - \frac{2GM}{r}, \hs{5.2} 
R_{\vf\thg\vf}^{\;\;\;\;\;\thg} = - \frac{2GM}{r}\, \sin^2 \thg. }
\ea
\label{a.3}
\ee

\section{Magnetic fields in Schwarzschild space-time \label{na2}}

We describe electromagnetic fields in Schwarzschild space-time in the frame of
a static distant observer associated with the Schwarzschild-Droste co-ordinate
system (\ref{a.0}). In the absence of electric components, any static magnetic 
fields can be derived in the radial gauge $A_r = 0$ from a vector potential 1-form 
\[
A = A_{\thg}\, d\thg + A_{\vf}\, d\vf.
\]
Imposing axial symmetry then implies that the components are functions of $(r, \thg)$
only. As a result the magnetic field strength $F_{ij} = \der_i A_j - \der_j A_i$ has 
components 
\be
F_{r\thg} = \der_r A_{\thg}, \hs{2} F_{r\vf} = \der_r A_{\vf}, \hs{2} F_{\thg\vf} = \der_{\thg} A_{\vf}.
\label{na2.1}
\ee 
The Maxwell equations for the magnetic field strength components (\ref{na2.1}) in the 
presence of a current density $j^i$ then reduce to 
\be
\der_j \lh \sqrt{-g} F^{ji} \rh = \der_j \lh \sqrt{-g}\, g^{jk} g^{il} F_{kl} \rh = \sqrt{-g}\, j^i.
\label{na2.3}
\ee
Written out separately for $i = (r, \thg, \vf)$:
\be
\ba{l}
\dsp{ \der_{\thg} \lh \sqrt{-g}\, g^{\thg\thg} g^{rr} F_{r\thg} \rh  = \sqrt{-g}\, j^r, }\\
 \\
\dsp{ \der_r \lh \sqrt{-g}\, g^{rr} g^{\thg\thg} F_{r\thg} \rh = \sqrt{-g}\, j^{\thg}, }\\
 \\
\dsp{ \der_r \lh \sqrt{-g}\, g^{rr} g^{\vf\vf} F_{r\vf} \rh + 
 \der_{\thg} \lh \sqrt{-g}\, g^{\thg\thg} g^{\vf\vf} F_{\thg\vf} \rh = \sqrt{-g}\, j^{\vf}. }
\ea
\label{na2.4}
\ee
In the absence of radial and polar currrents: $j^r = j^{\thg} = 0$, the first two 
equations imply that 
\be 
F_{r \thg} = \frac{\kg}{\sqrt{-g}}\, g_{rr} g_{\thg\thg},
\label{2.5}
\ee
with $\kg$ a constant; in Schwarzschild space-time this becomes
\be
F_{r\thg} = \frac{\kg}{\sin \thg}\, \frac{1}{1 - \frac{2GM}{r}}.
\label{2.6}
\ee
This solution is singular everywhere on the $z$-axis as well as on the horizon. 
Therefore we discard it by setting $A_{\thg} = 0$. 

Finally, evaluation of the third equation (\ref{na2.4}) yields the result \ct{petterson:1974}
\[
r^2 \der_r \left[ \lh 1 - \frac{2GM}{r} \rh \der_r A_{\vf} \right] + 
 \sin \thg \der_{\thg} \left[ \frac{1}{\sin \thg} \der_{\thg} A_{\vf} \right] = - r^4 \sin^2 \thg\, j^{\vf},
\]
which is defines the starting point eq.\ (\ref{n2.1}) for the discussion in sect.\ \ref{s2}.

The relation between the magnetic field strength components $F_{ij}$ and the dual 
axial vector field $B_i$ is defined in terms of the antisymmetric permutation tensor 
\be
E_{\mu\nu\rg\sg} = \sqrt{-g}\, \ve_{\mu\nu\rg\sg} = r^2 \sin \thg\, \ve_{\mu\nu\rg\sg},
\label{a.4}
\ee 
with $\ve_{\mu\nu\rg\sg} = +1$ for even permutations of $(\mu\nu\rg\sg) = (tr\thg\vf)$,
$\ve_{\mu\nu\rg\sg} = -1$ for odd permutations of $(\mu\nu\rg\sg) = (tr\thg\vf)$, and 
$\ve_{\mu\nu\rg\sg} = 0$ in all other cases. Its inverse is 
\be
E^{\mu\nu\rg\sg} = \frac{1}{\sqrt{-g}}\, \ve^{\mu\nu\rg\sg} = \frac{1}{r^2 \sin \thg}\, \ve^{\mu\nu\rg\sg},
\label{a.5}
\ee
with $\ve^{\mu\nu\rg\sg} = -1$ for even permutations of $(\mu\nu\rg\sg) = (tr\thg\vf)$
and $\ve_{\mu\nu\rg\sg} = +1$ for odd permutations of $(\mu\nu\rg\sg) = (tr\thg\vf)$. 
With these definitions the dual electromagnetic field strength is 
\be
\tilde{F}^{\mu\nu} = \frac{1}{2}\, E^{\mu\nu\rg\sg} F_{\rg\sg},
\label{a.6}
\ee
and the $B$-field components are defined by
\be
B_i = \tilde{F}_{0i} = g_{00} g_{ij} \tilde{F}^{0j} = - \frac{1}{2\sqrt{-g}}\, g_{00} g_{ij} \ve^{jkl} F_{kl}.
\label{a.7}
\ee
Therefore in Schwarzschild space-time
\be
B_r = \frac{1}{r^2 \sin \thg}\, F_{\thg\vf}, \hs{1} 
B_{\thg} = - \frac{1}{\sin \thg} \lh 1 - \frac{2GM}{r} \rh F_{r\vf}, \hs{1}
B_{\vf} = \lh 1 - \frac{2GM}{r} \rh F_{r\thg}.
\label{a.8}
\ee
The vanishing of the field-strength component $F_{r\thg}$ is seen to be equivalent 
to $B_{\vf} = 0$. 

\section{Angular dependence of magnetic fields \label{na3}}

The angular dependence of the magnetic fields discussed in this paper is expressed in terms 
of Legendre polynomials $P_l(\cos \thg)$ and the related functions $\Fg_l(\cos \thg)$ derived 
in section \ref{s2}. Here we give explicit expressions for the cases $l = 1,2,3,4$, or dipole, 
quadrupole, octopole and hexadecapole fields. Writing $x = \cos \thg$:
\be
\ba{l}
\dsp{ P_0(x) = 1, \hs{2} P_1(x) = x \hs{2} P_2(x) = \frac{1}{2} \lh 3 x^2 - 1 \rh, }\\
 \\
\dsp{ P_3(x) = \frac{1}{2} \lh 5 x^3 - 3 x \rh, \hs{2} P_4(x) = \frac{1}{8} \lh 35 x^4 - 30 x^2 + 3 \rh, }
\ea
\label{na3.0a}
\ee
and
\be
\ba{l}
\dsp{ \Fg_1(x) = - \lh x^2 - 1 \rh, \hs{2} \Fg_2(x) = - 3 \lh x^3 - x \rh, }\\
 \\
\dsp{ \Fg_3(x) = - \frac{3}{2} \lh 5 x^4 - 6 x^2 + 1 \rh, \hs{2} 
 \Fg_4(x) = - \frac{5}{2} \lh 7 x^5 - 10 x^3 + 3 x \rh. }
\ea
\label{na3.0b}
\ee

\np
\section{Radial dependence of magnetic fields \label{na4}}

In this appendix we provide details of the solutions for the radial dependence of the magnetic 
vector potential $A_{\vf}$. We first consider the series solution in inverse powers of $r$. 
Switching to the new variable $\eta = 2GM/r$ the first equation (\ref{n2.3}) takes the form
\be
\frac{d}{d\eta} \left[ \eta^2 \lh 1 - \eta \rh \frac{df_l}{d\eta}\right]  
 = \lh \eta^2 - \eta^3 \rh \frac{d^2f_l}{d\eta^2} + \lh 2\eta- 3 \eta^2 \rh \frac{df_l}{d\eta} = l(l+1)\, f_l.
\label{na3.1}
\ee
With the Ansatz
\be
f_l(\eta) = \sum_{n=0}^{\infty} c^{(l)}_n \eta^n, 
\label{2na3.2}
\ee
this becomes
\be
\ba{l}
\dsp{ l(l+1)\, c^{(l)}_0 + \left[ (l(l+1) - 2 \right] c^{(l)}_1 \eta + 
 \left[ \lh l(l+1) - 6 \rh c^{(l)}_2 + 3 c^{(l)}_1 \right] \eta^2 }\\
 \\
\dsp{ \hs{2.5} +\, \sum_{n=3}^{\infty} \left[ \lh l(l+1) - n(n+1) \rh c^{(l)}_n + 
 (n-1)(n+1)\, c^{(l)}_{n-1} \right] \eta^n = 0. }
\ea
\label{na3.3}
\ee
Since this is to hold for any $\eta$, it follows that $c_0^{(l)} = 0$ and for 
any $k \geq 1$:
\be
c^{(l)}_{l-k} = 0, \hs{1} c^{(l)}_{l+k} = \frac{(l+k)^2 - 1}{k(2l+k+1)}\,c^{(l)}_{l+k-1}.
\label{na3.4}
\ee
The upshot is, that $c_l^{(l)}$ is a free parameter, and higher coefficients are 
related to this parameter by 
\be
c^{(l)}_{l+k} = \frac{1}{k!}\, \prod_{m=1}^k \lh \frac{(l+m)^2 - 1}{2l+m+1} \rh c_l^{(l)}.
\label{na3.5}
\ee
The first few series then start out as
\be
\ba{lll}
f_1(\eta) & = & \dsp{c_1^{(1)} \eta \lh 1 + \frac{3}{4}\, \eta + \frac{3}{5}\, \eta^2 
 + \frac{1}{2}\, \eta^3 + ... \rh, }\\
 & & \\
f_2(\eta) & = & \dsp{ c_2^{(2)} \eta^2 \lh 1 + \frac{4}{3}\, \eta + \frac{10}{7}\, \eta^2 
 + \frac{10}{7}\, \eta^3 + ...\rh, }\\
 & & \\
f_3(\eta) & = & \dsp{ c_3^{(3)} \eta^3 \lh 1 + \frac{15}{8}\, \eta + \frac{5}{2}\, \eta^2  
 + \frac{35}{12}\, \eta^3 + ... \rh, }\\
 & & \\
f_4(\eta) & = & \dsp{ c_4^{(4)} \eta^4 \lh 1 + \frac{12}{5}\, \eta + \frac{42}{11}\, \eta^2 
 + \frac{56}{11}\, \eta^3 + ... \rh;}
\ea
\label{na3.6}
\ee
Note in particular that the dipole solution with $l = 1$ can be written in closed form as 
\be
f_1(\eta) = - \frac{3 c^{(1)}_1}{\eta^2} \left[ \eta + \frac{\eta^2}{2} + \ln \lh 1 - \eta \rh \right].
\label{na3.8.c}
\ee
{\em Convergence} 
Let us briefly consider the convergence of the expression for $f_l(r)$. The ratio test implies that
absolute convergence is implied if 
\be
\lim_{n \rightarrow \infty} \left| \frac{c^{(l)}_{n+1}}{c^{(l)}_n} \right| \eta < 1 \hs{1} 
 \Leftrightarrow \hs{1}  \frac{r}{2GM} > \lim_{k \rightarrow \infty} \left| \frac{c^{(l)}_{l+k}}{c^{(l)}_{l+k-1}} \right|. 
\label{na3.7}
\ee
Now eq.\ (\ref{na3.4}) shows that for $k \geq 1$
\be
\lim_{k \rightarrow \infty} \left| \frac{c^{(l)}_{l+k}}{c^{(l)}_{l+k-1}} \right| = \lim_{k \rightarrow \infty} 
 \lh \frac{(l+k)^2-1}{k(2l+k+1)} \rh = 1.
\label{na3.8}
\ee
Therefore the expression for $f_l(r)$ converges everywhere outside the black-hole horizon. 
As concerns zeros, note that all terms in parentheses in the series expansion for $f_l(r)$ are 
positive; therefore within the domain of convergence $r > 2 GM$ the function $f_l(r)$ has the 
same sign as $c_l^{(l)}$ everywhere; similarly $f'_l(r)0$ has the opposite sign everywhere 
outside the horizon, therefore depending on the sign of $c_l^{(l)}$ the function $f_l(r)$ is
monotonically decreasing or rising to zero with zero slope at $r \rightarrow \infty$; 
asymptotically for $r \rightarrow \infty$
\[
\frac{f_l(r)}{c_l^{(l)}} \downarrow 0, \hs{1} \mbox{and} \hs{1}
\frac{f'_l(r)}{c_l^{(l)}} \uparrow 0.
\]
As all eigenvalues (\ref{n2.8}) are different, we can construct an orthonormal basis for these 
functions by defining an inner product w.r.t.\ which the operator
\[
D = \frac{d}{d\eta} \left[ \eta^2 \lh 1 - \eta \rh \frac{d}{d\eta}\right], 
\]
is self-adjoint:
\be
\lh f_k, D f_l \rh = \lh D f_k, f_l \rh.
\label{na3.8a}
\ee
Given such an inner product, the functions 
\be
u_l(\eta) = \frac{f_l(\eta)}{\sqrt{(f_l,f_l)}}, \hs{2} \lh u_k, u_l \rh = \del_{kl},
\label{na3.8b}
\ee
form an orthonormal basis. A slight complication here arises, as the obvious definition
\[
\lh f_k, f_l \rh = \int_0^1 d\eta\, f_k(\eta) f_l(\eta),
\]
cannot apply, even though it satisfies the condition that $D$ is self-adjoint. The reason is 
that the functions $f_l(\eta)$ are not square integrable in this domain; therefore regularization 
must be applied. This issue does not affect the physical applications, as the multipole  
fields defining the basis used in (\ref{n2.9}) are well-defined everywhere in the domain 
$r > 2GM$.

In view of other possible applications we also briefly consider solutions with positive powers 
of $r$. We denote such solutions by $g_l(z)$, where by definition $z = 1/\eta = r/2GM$;
equation (\ref{n2.3}) then becomes:
\be
z^2 \frac{d}{dz} \left[ \lh 1 - \frac{1}{z} \rh \frac{dg_l}{dz} \right] = 
 z(z-1) \frac{d^2 g_l}{dz^2} + \frac{dg_l}{dz} = l \lh l+1 \rh g_l.
\label{na3.9}
\ee
We look for solutions 
\be
g_l(z) = \sum_{n=0}^{\infty} a^{(l)}_n z^n.
\label{na3.10}
\ee
After substitution in (\ref{na3.9}) this equation becomes
\be
\ba{l}
0 = l(l+1)\, a^{(l)}_0 - a^{(l)}_1 + l(l+1)\, a^{(l)}_1 z \\
 \\
\dsp{ \hs{2} +\, \sum_{n \geq 2} \left[ (n-1)(n+1) a^{(l)}_{n+1} + \lh l(l+1) - n (n-1) \rh a^{(l)}_n \right] z^n }.
\ea
\label{na3.11}
\ee
It follows that $a^{(l)}_0 = a^{(l)}_1 = 0$ and for $n \geq 2$
\be
a^{(l)}_{n+1} = \frac{n(n-1) - l (l+1)}{(n-1)(n+1)}\, a^{(l)}_n. 
\label{na3.12}
\ee
Then for all $k \geq 2$ also $a^{(l)}_{l + k} = 0$, and the solutions become polynomials 
of order $l+1$: 
\be
g_l(z) = \sum_{n=2}^{l+1} a^{(l)}_n z^n, 
\label{na3.14}
\ee
with $a^{(l)}_2$ a free parameter and all coefficients $a^{(l)}_{2+k}$ defined by
\be
a^{(l)}_{2+k} = \prod_{m=1}^k \lh \frac{m(m+1) - l(l+1)}{m(m+2)} \rh a^{(l)}_2, \hs{1} k = 1, ..., l-1.
\label{na3.15}
\ee
The first few polynomials are
\be
\ba{ll}
g_1(z) = a_2^{(1)}\, z^2, & \dsp{ g_2(z) = a_2^{(2)} z^2 \lh 1 - \frac{4}{3}\, z \rh, }\\
\\
\dsp{ g_3(z) = a_2^{(3)} z^2 \lh 1 - \frac{10}{3}\, z + \frac{5}{2}\, z^2 \rh, }&
\dsp{ g_4(z) = a_2^{(4)} z^2 \lh 1 - 6 z + \frac{21}{2}\, z^2 -  \frac{28}{5}\, z^3  \rh. } 
\ea
\label{na3.16}
\ee
All zeros of these polynomials are located within the horizon $z < 1$ and therefore 
not physically relevant. An orthonomal basis for these functions is defined by
\be
v_l(z) = \frac{g_l(z)}{\sqrt{(g_l, g_l)}}, 
\label{na3.17}
\ee
in terms of the inner product 
\be
(g_k, g_l) = \int_0^1 \frac{dz}{z^2}\, g_k(z) g_l(z),
\label{}
\ee
which is well-defined for the polynomial solutions. As all eigenvalues are non-degenerate
it follows that
\be
(v_k, v_l) = \del_{kl}.
\label{na3.18}
\ee

\section{Stability analysis for circular dipole orbits \label{na6}}

In section \ref{ns.6} we discussed the stability of circular orbits in dipole magnetic 
fields. This appendix presents the details of the analysis on which the results about 
stability are based. The input of this analysis is first the condition of existence of 
circular orbits, and second the periodicity of the fluctuations about these orbits, 
expressed by the reality of the frequencies of the transverse and planar deviations. 
In terms of the dimensionless variables (\ref{n6.x}) these conditions are summarized 
by eqs.\ (\ref{n6.1.a}), (\ref{n6.4}) and (\ref{n6.7}). The complication of analyzing these
conditions is, that they need to be considered separately in different domains of variables: 
$2 < x < 3$, $3 < x < 6$ and $x > 6$; and that they have to be discussed for two choices 
of sign in the relation between the angular velocity $\Og$ and equatorial magnetic field 
$y$. We first discuss the choice of positive sign:
\be
2 (x - 3) \Og = y \lh 1 + \sqrt{1 + \frac{4(x-3)}{x^2 y^2}} \rh.
\label{a6.0}
\ee
This relation imposes the inequality (\ref{n6.1.a}), which is trivially satisfied for $x \geq 3$; 
but for $2 < x < 3$ it requires a non-zero magnetic field: 
\be
y^2 > \frac{4(3-x)}{x^2}.
\label{a6.1}
\ee
Next, the inequality (\ref{n6.4}) for transverse stability becomes 
\be
\sqrt{1 + \frac{4(x-3)}{x^2y^2}} > - \lh 1 +  \frac{4(x-3)k(x)}{xh(x)} \rh.
\label{transversestability}
\ee
Again, it is satisfied for all $y$ in the range (\ref{a6.1}) if the r.h.s.\ is negative,
which happens if 
\be
xh(x) \geq 4(3-x) k(x).
\label{a6.2}
\ee
This is true automatically for $x \geq 3$; if $xh(x) < 4(3-x) k(x)$ the case $2 < x < 3$ has 
to be considered separately. Terms on both sides of the inequality can then be squared
and rearranged to give
\be
 y^2 > \frac{4(3-x)}{x^2} \frac{1}{1 - \lh \frac{4(3-x)k(x)}{xh(x)} -1 \rh^2}.
\label{a6.3}
\ee
Non-trivial solutions of this inequality implying transverse stability (\ref{transversestability})
exist only in the domain 
\be
0 < \frac{2(3-x) k(x)}{xh(x)} < 1,
\label{a6.4b}
\ee
or approximately $2.23 < x < 3$; see fig.\ F.1.

\bc
\scalebox{0.3}{\includegraphics{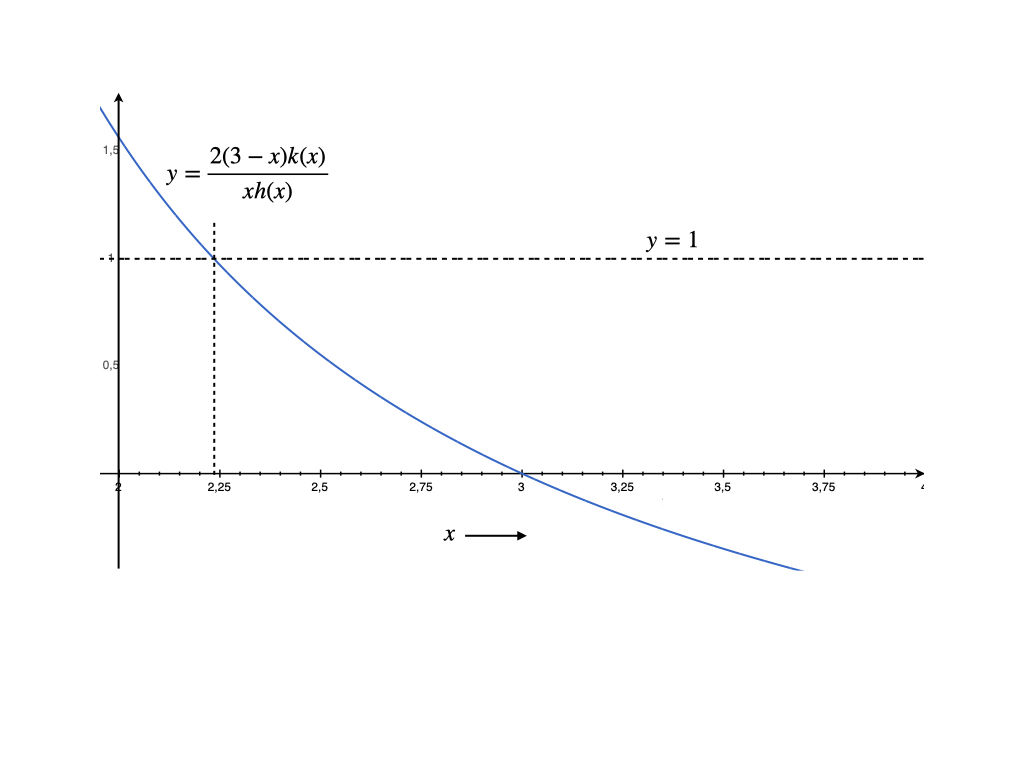}}
\vs{-7}

{\footnotesize \hs{1} Fig.\ F.1: Range of $x$ for which $ 2 < x < 3$ and $2(3-x)k(x)/xh(x) < 1$.}
\ec
In summary, either $x \geq 3$ for all values of $y^2$, or $2.23 < x < 3$ and $y^2$ in the range
\be
y^2 > \frac{4(3-x)}{x^2} \frac{1}{1 - \lh 1 - \frac{4(3-x)k(x)}{xh(x)} \rh^2} > \frac{4(3-x)}{x^2}.
\label{a6.5}
\ee
Finally it remains to consider the condition for planar stabilty, expressed by the inequality 
(\ref{n6.7}), which can be recast in the form
\be
\frac{(x-2)(x-3)(x-6)}{x^2 y^2} + (x-3)^2 - F(x) \lh 1 + \sqrt{1 + \frac{4(x-3)}{x^2y^2}} \rh \geq 0,
\label{a6n.1}
\ee
where
\be
F(x) \equiv \frac{1}{2} \lh x^2 - 4x + 6 + 2(x-2)(x-3) \frac{k(x)}{h(x)} \rh.
\label{a6n.2}
\ee
\bc
\scalebox{0.27}{\includegraphics{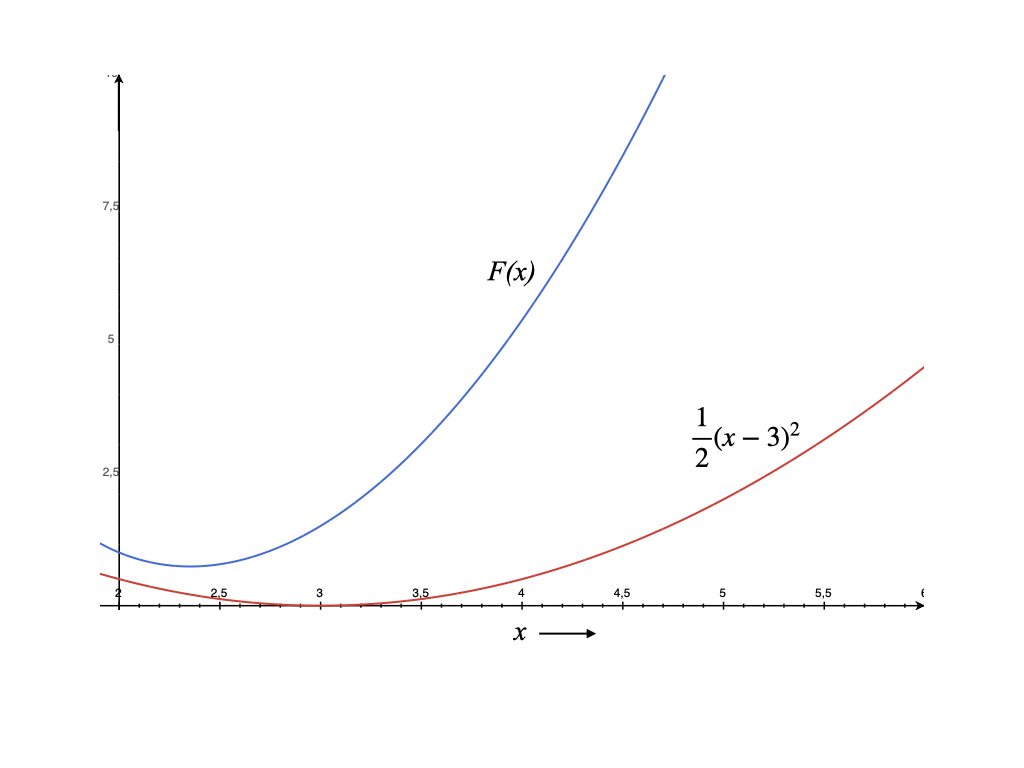}}
\vs{-5}

\footnotesize{Fig.\ F.2: The function $F(x)$ defined in eq.\ (\ref{a6n.2}).}
\ec
Fig.\ F.2 shows that in the domain $x > 2$ this function is everywhere positive.
For comparison we have also plotted the function $(x-3)^2/2$ to show that
\[
2 F(x) > \lh x - 3 \rh^2.
\]
This implies that the inequality (\ref{a6n.1}) is never satisfied in the limit $y^2 \rightarrow \infty$;
thus it imposes restrictions on the allowed values of the magnetic field as represented by $y$.
Defining the coefficients
\[
\ba{l}
\dsp{ A = (x-2)^2 (x-3)^2 (x- 6)^2, }\\
 \\
\dsp{ B = x^2 (x-3) \left[ (x-2)(x-6) \lh F(x) - (x - 3)^2 \rh + 2 F^2(x) \right], }\\
 \\
\dsp{ C = x^4 (x-3)^2 \lh F(x) - \frac{1}{2} (x-3)^2 \rh, }
\ea
\]
and considering the domain $x > 6$ where $(x - 2)(x-3)(x - 6) > 0$ and $(A, B, C) > 0$
everywhere, the condition of planar stability reduces to
\[
\frac{A}{y^4} - \frac{2B}{y^2} - 2C \geq 0,
\]
which is solved for
\be
y^2 \leq \frac{A}{B + \sqrt{B^2 + 2AC}}. 
\label{a6n.5}
\ee
For $3< x < 6$ it follows that $(x-2)(x-3)(x-6) < 0$; in that domain 
\[
\frac{(x-2)(x-3)(x-6)}{x^2 y^2} + (x-3)^2 - F(x) \lh 1 + \sqrt{ 1 + \frac{4(x-3)}{x^2y^2}} \rh < 0;
\]
thus the condition (\ref{a6n.1}) for planar stability has no solutions.
Finally, for $2 < x < 3$ the coefficient $A > 0$, but $B$ can become negative; however, as 
$\sqrt{B^2 + 2AC} > |B|$ there are still non-trivial solutions for $y$ in the domain defined 
by (\ref{a6n.5}), provided (\ref{a6.1}) holds:
\be
\frac{4(3 - x)}{x^2} \leq y^2 \leq \frac{A}{B + \sqrt{B^2 + 2AC}}.
\label{a6.10}
\ee
It remains to investigate the stability of circular orbits for which 
\be
2(x-3) \Og = y \lh 1 - \sqrt{1 + \frac{4(x-3)}{x^2 y^2}} \rh.
\label{a6.12}
\ee
Again we have to distinguish the domains $x > 3$ and $2 < x < 3$; in both cases
$y/\Og < 0$ and therefore the inequality (\ref{n6.2}) becomes non-trivial:
\be
0 > \frac{y}{\Og} = \frac{2(x-3)}{1 - \sqrt{1 + \frac{4(x-3)}{x^2y^2}}}  > - \frac{xh(x)}{2k(x)}.
\label{a6.13}
\ee
For $x > 3$ this imposes the explicit constraint
\be
0 < y^2 < \frac{4(x-3)}{x^2} \frac{1}{\lh 1 + \frac{4(x-3) k(x)}{xh(x)} \rh^2 - 1}.
\label{a6.14}
\ee
For $2 < x < 3$ we get 
\be
y^2 < \frac{4(3-x)}{x^2} \frac{1}{1 - \lh \frac{4(3-x) k(x)}{xh(x)} - 1 \rh^2},
\label{a6.15}
\ee
with $x$ in the range (\ref{a6.4b}); see fig.\ F.1.

Finally, we have to impose the inequality (\ref{n6.7}), which in this case becomes 
\be
\frac{(x-2)(x-3)(x-6)}{x^2y^2} + (x-3)^2 + F(x) \lh \sqrt{1 + \frac{4(x-3)}{x^2y^2}} - 1 \rh \geq 0.
\label{a6.16}
\ee
For $x > 6$ it is always satisfied. Next, for $3 < x < 6$ it still holds in the limit 
$y^2 \rightarrow \infty$, but there is a lower threshold 
\be
y^2 \geq \frac{A}{B + \sqrt{B^2 + 2AC}}.
\label{a6.17}
\ee
In the remaining domain $2 < x < 3$ the situation is similar, but there is the additional 
condition (\ref{a6.1}); therefore the condition for stability of circular orbits in the plane 
becomes 
\be
y^2 \geq \mbox{max} \lh \frac{A}{B + \sqrt{B^2 + 2AC}}, \frac{4(3-x)}{x^2} \rh. 
\label{a6.18}
\ee

\section{Weak field limit of Kerr geometry \label{na5}}

The weak-field approximation of Kerr geometry, deviating from Minkowski space-time only
by terms linear in $G$, is described by the line element (\ref{6.1}). Based on this, the 
expressions for the components of the connection to first order in $G$ read
\be
\ba{lll}
\Gam_{rr}^{\;\;\;r} = - \frac{GM}{r^2}, & 
\Gam_{\thg\thg}^{\;\;\;r} = - r \lh 1 - \frac{2GM}{r} \rh, & \Gam_{r\thg}^{\;\;\;\thg} = \frac{1}{r}, \\
 & & \\
\Gam_{\vf\vf}^{\;\;\;r} = - r \sin^2 \thg \lh 1 - \frac{2GM}{r} \rh, & \Gam_{r\vf}^{\;\;\;\vf} = \frac{1}{r}, & \\
 & \\
\Gam_{tt}^{\;\;\;r} = \frac{GM}{r^2}, & \Gam_{rt}^{\;\;\;t} = \frac{GM}{r^2}, & \\
 & \\
\Gam_{\vf\vf}^{\;\;\;\thg} = - \sin \thg \cos \thg, & \Gam_{\thg\vf}^{\;\;\;\vf} = \frac{\cos \thg}{\sin \thg}, & \\
 & \\
\Gam_{r\vf}^{\;\;\;t} = - \frac{3GJ}{r^2}\, \sin^2 \thg, & \Gam_{rt}^{\;\;\;\vf} = \frac{GJ}{r^4}, 
 & \Gam_{t\vf}^{\;\;\;r} = - \frac{GJ}{r^2}\, \sin^2 \thg, \\
 & \\
\Gam_{\thg\vf}^{\;\;\;t} = \frac{GJ}{r}\, \sin \thg \cos \thg, & 
 \Gam_{t \thg}^{\;\;\;\vf} = - \frac{GJ}{r^3}\, \frac{\cos \thg}{\sin \thg}, & 
 \Gam_{t\vf}^{\;\;\;\thg} = \frac{GJ}{r^3}\, \sin \thg \cos \thg. 
\ea
\label{a3.1}
\ee
Non-zero $\thg$-components of the Riemann tensor: 
\be
\ba{ll}
\dsp{ R_{t\thg t}^{\;\;\;\;\;\thg} = - \frac{GM}{r^3}, }& 
\dsp{ R_{r\thg r}^{\;\;\;\;\;\thg} = \frac{GM}{r^3}, }\\
 & \\
\dsp{ R_{\vf\thg\vf}^{\;\;\;\;\;\thg} = - \frac{2GM}{r}\, \sin^2 \thg, }&   
\dsp{ R_{t\thg\vf}^{\;\;\;\;\;\thg} = \frac{2GJ}{r^3}\, \sin^2 \thg, }\\
\ea
\label{a3.2}
\ee

\np

\end{document}